# Insulating-to-Conducting Behavior and Band Profile Across the La$_{0.9}$Ba$_{0.1}$MnO$_3$/Nb:SrTiO$_3$ Epitaxial Interface

Weiwei Li*, Josée E. Kleibeuker, Rui Wu, Kelvin H. L. Zhang, Chao Yun, Judith L. MacManus-Driscoll*

*Department of Materials Science & Metallurgy, University of Cambridge, 27 Charles Babbage Road, Cambridge, CB3 0FS, United Kingdom*

E-mail: wl337@cam.ac.uk and jld35@cam.ac.uk

Abstract

La$_{0.9}$Ba$_{0.1}$MnO$_3$ is a ferromagnetic insulator in its bulk form, but exhibits metallicity in thin film form. It has a wide potential in a range of spintronic-related applications, and hence it is critical to understand thickness-dependent electronic structure in thin films as well as substrate/film interface effects. Here, using electrical and *in-situ* photoemission spectroscopy measurements, we report the electronic structure and interface band profile of high-quality layer-by-layer-grown La$_{0.9}$Ba$_{0.1}$MnO$_3$ on single crystal Nb:SrTiO$_3$ substrates. A transition from insulating-to-conducting was observed with increasing La$_{0.9}$Ba$_{0.1}$MnO$_3$ thickness, which was explained by the determined interface band diagram of La$_{0.9}$Ba$_{0.1}$MnO$_3$/Nb: SrTiO$_3$, where a type II heterojunction was formed.

Perovskite transition metal oxides have stimulated intense research activities owing to their strongly correlated electronic and physical properties arising from the interplay of lattice, charge, spin and orbital degrees of freedom. In recent years, with the perspective toward multifunctional devices, increased attention has turned to oxide heterostructures of these materials. The mixed-valence manganite, $La_{1-x}A_xMnO_3$ (A=Ca, Sr, and Ba), is one of the most studied strongly correlated systems because of its wide range of magnetic and conducting properties such as the colossal magnetoresistance (CMR), half metallicity, and ferromagnetic insulating behavior.[1-3] An extensive number of studies on the behavior of manganite oxide heterostructures have demonstrated that novel functionalities and new electronic phases can be generated by interface effects, such as orbital reconstruction,[4,5] charge transfer,[6,7] and electric field effects.[8-12]

Similar to the case of traditional semiconductors, an in-depth understanding of the interfacial electronic structure at the oxide interfaces is of vital importance for a better control of their intriguing properties and for the engineering of oxide electronic devices. Generally, the macroscopic properties of the devices are crucially dependent on the energy band alignment at the interface. The study of band alignment in oxide heterostructures provides important information about the current transport, potential distribution, and quantum carrier confinement at the interface.[13] Recently, interface band profiles of $La_{0.6}Sr_{0.4}MnO_3$/Nb-doped $SrTiO_3$ (Nb:STO),[14] $La_{2/3}Sr_{1/3}MnO_3$/Nb:STO,[15] $La_{0.7}Sr_{0.3}MnO_3$/Nb:STO,[16] $La_{2/3}Ca_{1/3}MnO_3$/Nb:STO,[17] and $La_{0.7}Ca_{0.3}MnO_3$/Nb:STO[18] heterostructures have been widely investigated. In particular, it has been shown that the Schottky barrier height (SBH) between polar $La_{0.6}Sr_{0.4}MnO_3$ (or $La_{0.7}Sr_{0.3}MnO_3$) and non-polar Nb:STO can be modified by inducing an interfacial dipole to screen the polar discontinuity at the interface *via* changing the interfacial termination layer.[14,16]

Compared to $La_{1-x}A_xMnO_3$ (A = Ca and Sr), $La_{1-x}Ba_xMnO_3$ shows a higher ferromagnetic (FM) Curie temperature ($T_C$) for lower doping levels.[19,20] Of particular interest is $La_{0.9}Ba_{0.1}MnO_3$ (LBMO), which in bulk is a ferromagnetic-insulator with an orthorhombic structure.[21,22] The combined magnetic and insulating properties give LBMO strong potential for spintronic devices, e.g. as a spin filter in tunnel magnetoresistance devices[23] or a component in a magnetoelectric composite system[24]. However, while bulk LBMO is insulating with a FM $T_C$ of 185 K, thin films of LBMO grown on (001)-oriented $SrTiO_3$ (STO) are typically metallic with higher FM $T_C$s.[8,22] The modified properties have been attributed to orbital reconstruction caused by the in-plane tensile strain from the STO substrate.[25] Epitaxial strain also plays a role in tuning properties of films with thickness well beyond the interfacial regime.[22]

As far as tuning of LBMO properties goes, it is promising that electrical modulation of double exchange ferromagnetism has been realized in LBMO/Nb:STO *p-n* heterojunctions.[8] However, the precise origins of the effect were not studied in detail. A comprehensive understanding of the origins of this effect may enable the interface properties of LBMO to be precisely electrically tuned, potentially enabling new spintronic devices to emerge. The starting point should be the determination of both the electronic structure of LBMO/Nb:STO films as a function of thickness and the interfacial band alignment. High-quality LBMO films are necessary for such studies.

In this study, highly controlled, ultrathin (5 – 40 unit cells), layer-by-layer-grown LBMO/Nb:STO heterojunctions were studied by *in-situ* x-ray photoelectron spectroscopy (XPS), current density-voltage (*J-V*), capacitance-voltage (*C-V*), and transport measurements. Films were fabricated on $TiO_2$ terminated (001)-oriented Nb:STO substrates. STO substrate with Nb doping level of 0.5 wt% were adopted to prevent charging effects during XPS measurements and also serve as a bottom electrode for the electrical measurements. The depositions were carried out at a substrate temperature of 780 $^o$C in an oxygen partial pressure of 150 mTorr with a repetition rate of 1 Hz and a laser fluence of 2 J·cm$^{-2}$. The growth dynamics were investigated by real-time monitoring the intensity variations of various features in the reflection high-energy electron diffraction (RHEED) patterns, allowing precise control of the thickness at the unit cell level. To ensure the oxygen stoichiometry, the films were *in-situ* annealed at 700 $^o$C under an oxygen pressure of 760 Torr before cooling to room temperature at a rate of 10 $^o$C/min. After cooling, the LBMO films were *in-situ* transferred to the photoemission chamber without breaking the vacuum (<10$^{-10}$ Torr), preventing the sample from surface contamination. Photoemission spectroscopy (PES) was undertaken using a monochromatic Al Kα x-ray source (*hv*= 1486.6 eV) using a SPECS PHOIBOS 150 electron energy analyzer with a total energy resolution of 0.5 eV. The Fermi level of the films was calibrated using a polycrystalline Au foil. Work function measurements were determined in a separate UHV system consisting of a monochromatic Al K$\alpha_1$ x-ray source together with a five channeltron PSP electron energy analyzer. The PSP spectrometer was calibrated using a clean silver foil and the measurements made at an overall resolution of 0.3 eV. The work function was determined by measuring the XPS secondary electron cutoff with the sample at a bias voltage of -10 V. The overall precision of the measurement was ± 0.05 eV. Film structures and surface morphologies were examined by x-ray diffraction (XRD, high resolution Panalytical Empyrean vertical diffractometer, Cu K$\alpha$ radiation) and atomic force microscopy (AFM), respectively. A Keithley 2400 source measure unit and an Agilent 4294A precision impedance analyser were used to measure the current-voltage and the capacitance-voltage characteristics of Pt/LBMO/Nb:STO at room temperature. The top Pt electrodes with a diameter of 200 μm were deposited by sputtering. Bias polarity was defined as positive or negative based on positive or negative voltage applied to the Pt electrodes. Electrical resistance was measured using a physical property measurement system (PPMS 14T, Quantum Design).

Figure 1 shows the RHEED specular intensities recorded during the growth of LBMO on TiO$_2$-terminated Nb:STO substrates. Clear RHEED intensity oscillations were observed during the entire growth of all samples, indicating controlled layer-by-layer growth of individual unit cells. The insets show the RHEED diffraction patterns of all samples captured after the growth. The sharp two-dimensional spots are present and lying on concentric Laue circles, implying true reflective diffraction from a smooth surface without the formation of 3D islands. Atomically smooth film surfaces with a well-defined terrace structure and one unit cell steps (~0.4 nm) were confirmed by *ex-* *situ* atomic force microscopy, as shown in Fig. 2 (a). Figure 2 (b) shows a typical out-of-plane XRD θ-2θ scan of 40 unit cells LBMO film grown on the Nb:STO substrate. The arrow indicates the $(002)_{pc}$ peak of LBMO with Kiessig fringes, suggesting the LBMO film has long range periodicity and a $(00l)$ orientation without the presence of any secondary phases or impurities. In order to determine in-plane lattice matching, reciprocal space maps (RSMs) of a 40 unit cell LBMO film were measured and are shown in Fig. 2(c). It is clearly seen that the LBMO film and the Nb:STO substrate have the same in-plane lattice parameters, demonstrating that the film is fully strained to the substrate. In addition, there is no difference in the (103), (013), (-103), and (0-13) Bragg peak positions, revealing that the 40 unit cells LBMO films grown on Nb:STO have a tetragonal distortion.

Figure 3 shows the current density versus voltage (*J-V*) characteristics of a Pt/LBMO(40 unit cells)/Nb:STO heterojunction at room temperature. By applying both reverse and forward voltages, distinct rectifying behavior was observed. An Ohmic contact between the electrode and the film was observed from the linear current-voltage curves. The rectifying ratio measured at ±2 V was 400, which is of the same order as reported values from other manganite oxides heterojunctions.[16,26,27] A turn on voltage of 0.78 V, a breakdown voltage of -2 V, and a forward current density of 30 A/cm$^2$ at 2 V were determined. Assuming transport by thermionic emission, the forward *J-V* characteristic (as shown in the inset of Fig. 3a) of a heterojunction can be described by the exponential relation[15, 28-30]

$$J \propto \exp\left(\frac{qV}{nk_BT}\right) \quad\quad\quad (1)$$

where $q$ is the charge of the electron, $k_B$ is the Boltzmann constant, $T$ is the temperature, and $n$ is the ideality factor. At the low bias, $n$ is estimated to be 1.8. This value is very close to 2 and indicates that thermionic emission dominates transport in this range. At the voltage range of 0.78-1.15 V, the calculated ideality factor was ~ 3.6, which might be attributed to the presence of additional interface states or the thermionic-field emission.[31,32]

The reverse bias junction capacitance is related to the applied voltage by[28]

$$\frac{1}{C^2} \propto (V_{bi} - V) \quad\quad\quad (2)$$

where $C$ is the reverse bias capacitance per unit area, $V$ is the reverse bias voltage, and $V_{bi}$ is the built-in voltage. We can directly derive the built-in potential ($qV_{bi}$) from the intercept point on the voltage axis, which corresponds to the energy difference in the Fermi levels ($E_f$) of the contacting materials. The deduced built-in potential value is 0.7 eV (as shown in Fig. 3(b)). However, the $J$-$V$ measurements usually underestimate the barrier height and give little information on the interface electronic structure.[16]

*In-situ* high resolution XPS was used to further determine the valence band offset and built-in potential at the interface. As shown in Fig. 4 (a-c), Sr 3d and Ti $2p_{3/2}$ from Nb:STO, and La 4d core-levels shift towards lower binding energy as the LBMO overlayer thickness increased. This indicates that the energy band of Nb:STO bends upwards and the energy band of LBMO bends downwards near the interface in the LBMO/Nb:STO heterostructures. The band bending on both the Nb:STO and LBMO sides caused by the deposition of LBMO were directly determined from the binding energy shifts of Sr 3d, Ti 2p$_{3/2}$, and La 4d core-level peaks, respectively. It should be noted that the measurements reflect the Sr 3d and Ti 2p$_{3/2}$ core-levels positions in the interface region due to the short electron escape depth of 3-5 nm for the XPS technique. A pure Nb:STO substrate and 5 unit cells LBMO film were used as reference samples to obtain the built-in potential. The binding energy shifts of Sr 3d$_{5/2}$, Ti 2p$_{3/2}$, and La 4d$_{5/2}$ core-level peaks are summarized in Fig. 4(d). Judging from the saturation levels of the peak shifts, the energy shifts can be estimated to be 0.8 ± 0.05 eV and 0.15 ± 0.05 eV for Nb:STO and LBMO, respectively. In other words, the built-in potential in the LBMO/Nb:STO heterojunctions is 0.95 ± 0.1 eV. Within the precision level of the measurement, this value is close to the value evaluated from the $1/C^2$ versus $V$ plot.

To obtain information about the band offset between LBMO and Nb:STO, the valence band offsets (VBOs, $\Delta E_V$) were calculated based on the method developed by Kraut $et$ $al$.[33,34]

$$\Delta E_V = (E_{CL} - E_V)^{thick\ film} - (E_{CL} - E_V)^{bare\ substrate}$$

$$-(E_{CL}^{thin\ film} - E_{CL}^{substrate}) \qquad (3)$$

Where $E_{CL}$ is a chosen core-level energy from an element in each material (not common in both the film and substrate) and $E_V$ is valence band energy. Here, for a thick film, a 40 unit cell LBMO film was used and, for the thin film, a 5 unit cell LBMO film was used. The Mn 2p$_{3/2}$ peak from LBMO, Ti 2p$_{3/2}$ peak from Nb:STO, and their valence bands (VBs) were measured to calculate the VBOs (Fig. 5(a)). The valence-band maximum (VBM, $E_V$) values were determined by linear extrapolation of the leading edge of the VB region to the extended baseline of the VB spectra in order to account for the instrumental broadening. The values of VBM are estimated to be -0.06(5) eV for 40 unit cells of LBMO, 0.42(5) eV for 5 unit cells of LBMO, and 3.18(5) eV for the Nb:STO substrate.

The inset of Fig. 5(a) shows an enlarged view of the VB near $E_f$. There is no intensity at $E_f$ for the 5 unit cell LBMO, whereas a small quantity of intensity appears at $E_f$ for the 40 unit cell LBMO. This indicates that 5 unit cell LBMO films are insulating and that 40 unit cell LBMO films are metallic. These results were confirmed by transport measurements (as shown in Fig. 5(c)). The observed insulating-to-metallic transition behavior in the 40 unit cell LBMO films is in agreement with 15 - 50 nm LBMO films with the same composition grown on STO substrates from previous literature and can be explained by the modification of Mn-O-Mn bond angles and bond lengths or $e_g$ orbital occupation.[8,22] The core-level peak position was defined to be the center of the peak width at half of the peak height. This procedure made it unnecessary to resolve the mixed valence state of Mn $2p_{3/2}$ core-levels for obtaining the high-precision peak position. As shown in Fig. 5(a), the obtained XPS peak positions are 642.11 eV and 642.20 eV for Mn $2p_{3/2}$ in the 40 and 5 unit cell LBMO films, respectively. The XPS peak positions of the Ti $2p_{3/2}$ core-levels are estimated to be 458.70 eV for the 5 unit cell films and 459.43 eV for the Nb:STO substrate. Thus, the calculated VBO is $\Delta E_V$ = 2.42 ± 0.1 eV. In addition, the work functions of a 40 unit cell LBMO film and a $TiO_2$-terminated Nb:STO substrate were measured using XPS by taking the difference between the x-ray source energy (1486.6 eV) and the secondary electron cutoffs (Fig.

5(b)) and are determined to be 5.0 eV and 3.9 eV, respectively. The measured work function of the Nb:STO substrate agree well with previous reported data.[35]

The band alignment at the LBMO/Nb:STO interface can be deduced from the *J-V* curves and XPS experiments. Figure 6 (a) shows the equilibrium band diagram of LBMO/Nb:STO heterojunctions after the alignment of $E_f$. It is a typical staggered energy band diagram (a type II heterojunction). In this diagram, an ideal interface is assumed where the defect traps at the LBMO/Nb:STO interface are ignored and the Nb:STO is a non-degenerate semiconductor.[36,37] The conduction band offsets (CBOs) can be calculated by $\Delta E_C = \Phi_{LBMO} - \Phi_{Nb:STO} - qV_{bi} = 0.08 \pm 0.05 \ eV$. The barrier height seen by electrons and holes in a *p-n* heterojunction should take account of the band offsets.[38] Therefore, the barrier height of 1.03 eV and 3.37 eV is obtained for electrons and holes in the LBMO/Nb:STO heterojunctions, respectively. Clearly, the barrier height for holes is significantly larger than that of electrons, thus electrons should be the main carriers under forward bias. In this case, the ideal turn-on voltage is 1.03 V, which is slightly larger than the value determined from the *J-V* curve (see Fig. 3(a)). This may be caused by the formation of an interface dipole. Assuming a fully ionic charge assignment using the nominal valence for each layer, the present interface has a polar discontinuity at the interface having a charge density of $-0.9q(MnO_2)$ $/+0.9q(La_{0.9}Ba_{0.1}O)$ $/0q(TiO_2)$ $/0q$ (SrO) for an *n*-type interface.

Based on the above discussion, a schematic density of states for the LBMO/Nb:STO heterojunction is proposed and shown in Fig. 6(b). The $E_f$ of homogeneous Nb:STO is located slightly below the bottom of the Ti 3d conduction band. For the LBMO films, there are two different regions. In region 1, the $e_g$ state of Mn 3d orbitals is slightly below $E_f$, which explains the insulating properties of the 5 unit cell LBMO films. In region 2, there is a small quantity of $e_g$ state of Mn 3d orbitals on $E_f$, which explains the metallic nature of the 40 unit cell LBMO films. The mechanism for the insulating behavior shown in the 5 unit cells film is consistent with the electronic reconstruction and/or modification of Mn-O-Mn bond angles and bond lengths close to the interface as occurs in ultrathin $La_{0.67}Sr_{0.33}MnO_3$ films grown on STO.[4,39] For ultrathin $La_{1-x}Sr_xMnO_3$ (x = 0.3-0.4) films, the insulating behavior has been shown below 7-8 unit cells.[40,41] Also, it has been reported that the $La_{1-x}Sr_xMnO_3$ (x = 0.3-0.4)/Nb:STO interfaces form Schottky junctions.[14-16] However, to the best of our knowledge, the insulating to conducting transition has not been observed in the band diagram. This is likely because the hole concentration of $La_{1-x}Sr_xMnO_3$ (x = 0.3-0.4) is much higher than in our LBMO films, and so the depletion width at the interface of $La_{1-x}Sr_xMnO_3$ (x = 0.3-0.4)/Nb:STO is much smaller.

   In summary, the electrical properties, electronic structure and interface band alignment of different thickness (5 – 40 unit cells) LBMO films grown on Nb:STO substrates were investigated. It was found that there is a transition from insulating-to-conducting behavior as the LBMO thickness increases. To explain the transition, the electronic structure and interface band alignment of the LBMO/Nb:STO heterojunctions was determined. A type II heterojunction was formed at the interface of LBMO/Nb:STO. The present results emphasize the importance of the interface between the film and the supporting substrate, with drastic effects on the electronic structure and transport properties. More broadly, the results show that determining the electronic structure at interface regions in correlated oxide thin films is crucial for providing a clear direction for oxide electronic device design.

**Acknowledgements**

This work was supported by EPSRC grant EP/K035282/1, EPSRC grant EP/N004272/1, and the Isaac Newton Trust (Minute 13.38(k)). K.H.L.Z. acknowledges the funding support from Herchel Smith Postdoctoral Fellowship by University of Cambridge.

# References

1. A.-M. Haghiri-Gosent and J.-P. Renard, J. Phys. D: Appl. Phys. **36**, R127 (2003).

2. H. Yamada, Y. Ogawa, Y. Ishii, H. Sato, M. Kawasaki, H. Akoh and Y. Tokura, Science **305**, 646 (2004).

3. Y. K. Liu, Y. W. Yin and X. G. Li, Chin. Phys. B 22, 087502 (2013).

4. A. Tebano, C. Aruta, S. Sanna, P. G. Medaglia, G. Balestrino, A. A. Sidorenko, R. De Renzi, G. Ghiringhelli, L. Braicovich, V, Bisogni and N. B. Brookes, Phys. Rev. Lett. **100**, 137401 (2008).

5. P. Yu, J.-S. Lee, S. Okamoto, M.D. Rossell, M. Huijben, C.-H. Yang, Q. He, J. X. Zhang, S. Y. Yang, M. J. Lee, Q. M. Ramasse, R. Erni, Y.-H. Chu, D. A. Arena, C.-C. Kao, L. W. Martin and R. Ramesh, Phys. Rev. Lett. **105**, 027201 (2010).

6. J. Chakhalian, J. W. Freeland, H.-U. Habermeier, G. Cristiani, G. Khaliulin, M. van Veenendaal and B. Keimer, Science **318**, 1114 (2007).

7. T. Y. Chien, L. F. Kourkoutis, J. Chakhalian, B. Gray, M. Kareev, N. P. Guisinger, D. A. Muller and J. W. Freeland, Nat. Commun. **4**, 2336 (2013).

8. H. Tanaka, J. Zhang and T. Kawai, Phys. Rev. Lett. **88**, 027204 (2001).

9. X. Hong, A. Posadas and C. H. Ahn, Phys. Rev. B **68**, 134415 (2003).

10. B. Cui, C. Song, H. Mao, Y. Yan, F. Li, S. Gao, J. Peng, F. Zeng and F. Pan, Adv. Funct. Mater. **26**, 753 (2016).

11. A. Herklotz, E.-J. Guo, A. T. Wong, T. L. Meyer, S. Dai, T. Z. Ward, H. N. Lee and M. R. Fitzsimmons, Nano Lett. **17**, 1665 (2017).

12. M. Gajek, M. Bibes, S. Fusil, K. Bouzehouane, J. Fontcuberta, A. Barthélémy and A. Fert, Nat. Mater. 6, 296 (2007).

13. S. M. Sze and K. K. Ng, *Physics of Semiconductor Devices* (Wiley-Interscience, 2007).

14. M. Minohara, R. Yasuhara, H. Kumigashira and M. Oshima, Phys. Rev. B **81**, 235322 (2010).

15. K. G. Rana, S. Parui and T. Banerjee, Phys. Rev. B **87**, 085116 (2013).

16. Y. Hikita, M. Nishikawa, T. Yajima and H. Y. Hwang, Phys. Rev. B **79**, 073101 (2009).

17. T. Y. Chien, J. Liu, J. Chakhalian, N. P. Guisinger and J. W. Freeland, Phys. Rev. B **82**, 041101 (2010).

18. F. A. Cuellar, G. Sanchez-Santolino, M. Varela, M. Clement, E. Iborra, Z. Sefrioui, J. Santamaria and C. Leon, Phys. Rev. B **85**, 245122 (2012).

19. R. von Helmolt, J. Wecker, B. Holzapfel, L. Schultz and K. Sam-wer, Phys. Rev. Lett. **71**, 2331 (1993).

20. H. L. Ju, Y. S. Nam, J. E. Lee and H. S. Shin, J. Magn. Magn. Mater. 2000, 219, 1.

21. B. Dabrowski, K. Rogacki, X. Xiong, P. W. Klamut, R. Dybzinski, J. Shaffer and J. D. Jogensen, Phys. Rev. B **58**, 2716 (1998).

22. J. Zhang, H. Tanaka, T. Kanki, J.-H. Choi and T. Kawai, Phys. Rev. B **64**, 184404 (2001).

23. M. Gajek, M. Bibes, A Bathélémy, K. Bouzehouane, S. Fusil, M. Varela, J. Fontcuberta and A. Fert, Phys. Rev. B **72**, 020406(R) (2005).

24. T. Fix, E.-M. Choi, J. W. A. Robinson, S. B. Lee, A. Chen, B. Prasad, H. Wang, M. G. Blamire and J. L. MacManus-Driscoll, Nano. Lett. **13**, 5886 (2013).

25. T. Kanki, H. Tanaka and T. Kawai, Phys. Rev. B **64**, 224418 (2001).

26. K.-J. Jin, H.-B. Lu, Q.-L. Zhou, K. Zhao, B.-L. Cheng, Z.-H. Chen, Y.-L. Zhou and G.-Z. Yang, Phys. Rev. B **71**, 184428 (2005).

27. M. Minohara, Y. Furnkawa, R. Yasuhara, H. Kumigashira and M. Oshima, Appl. Phys. Lett. **94**, 242106 (2009).

28. M. Sze, *Physics of Semiconductor Devices*, 2nd Ed. (Wiley, New York, 1969).

29. E. H. Rhoderick and R. H. William, *Metal-Semiconductor Contacts*, 2nd Ed. (Clarendon, Oxford, 1988).

30. A. Sawa, T. Fujii, M. Kawasaki and Y. Tokura, *Appl. Phys. Lett.* 2005, 86, 112508.

31. A. Ruotolo, C. Y. Lam, W. F. Cheng, K. H. Wong and C. W. Leung, Phys. Rev. B **76**, 075122 (2007).

32. T. Susaki, N. Nakagawa and H. Y. Hwang, Phys. Rev. B **75**, 104409 (2007).

33. E. A. Kraut, R. W. Grant, J. R. Waldrop and S. P. Kowalczyk, Phys. Rev. Lett. **44**, 1620 (1980).

34. E. A. Kraut, R. W. Grant, J. R. Waldrop and S. P. Kowalczyk, Phys. Rev. B **28**, 1965 (1983).

35. J. Robertson, J. Vac. Sci. Technol. B **18**, 1785 (2000).

36. S. M. Sze, *Semiconductor Devices: Physics and Technology*, 2nd Ed. (Wiley, New York, 2002).

37. Y. Watanabe, Phys. Rev. B **57**, R5563(R) (1998).

38. D. A. Neamen, *Semiconductor Physics and Devices: Basic Principles*, 3rd Ed. (McGraw-Hill, New York, 2003).

39. H. Boschker, J. Kautz, E. P. Houwman, W. Siemons, D. H. Blank, M. Huijben, G. Koster, A. Vailionis and G. Rijnders Phys. Rev. Lett. **109**, 157207 (2012).

40. M. Huijben, L. W. Martin, Y.-H. Chu, M. B. Holcomb, P. Yu, G. Rijnders, D. H. A. Blank and R. Ramesh Phys. Rev. B **78**, 094413 (2008).

41. R. Peng, H. C. Xu, M. Xia, J. F. Zhao, X. Xie, D. F. Xu, B. P. Xie and D. L. Feng Appl. Phys. Lett. **104**, 081606 (2014).

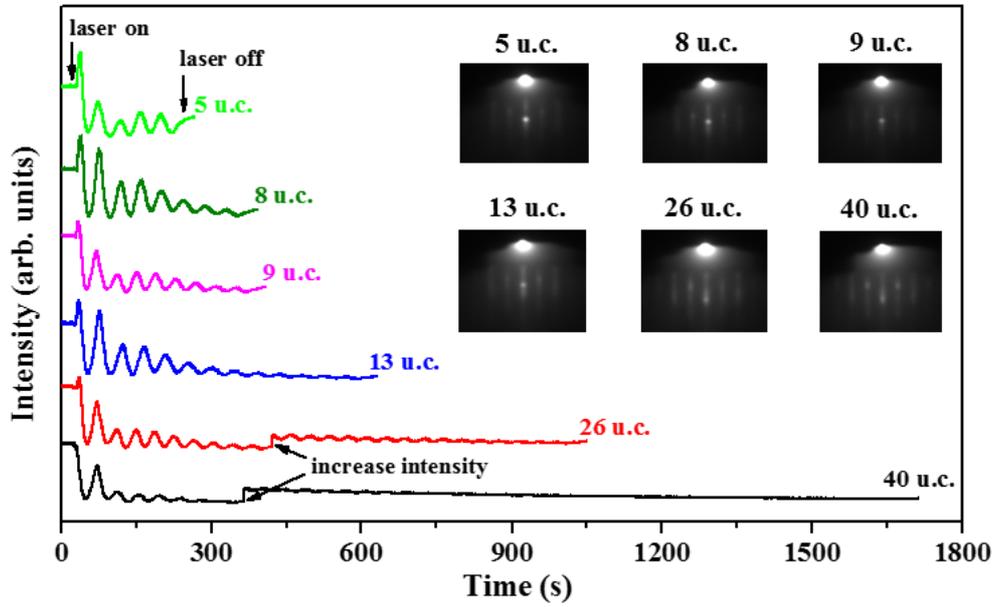

Figure 1. RHEED oscillations recorded during the growth of $La_{0.9}Ba_{0.1}MnO_3$ films with a thickness from 5 to 40 unit cells. The insets display the RHEED diffraction patterns after the growth.

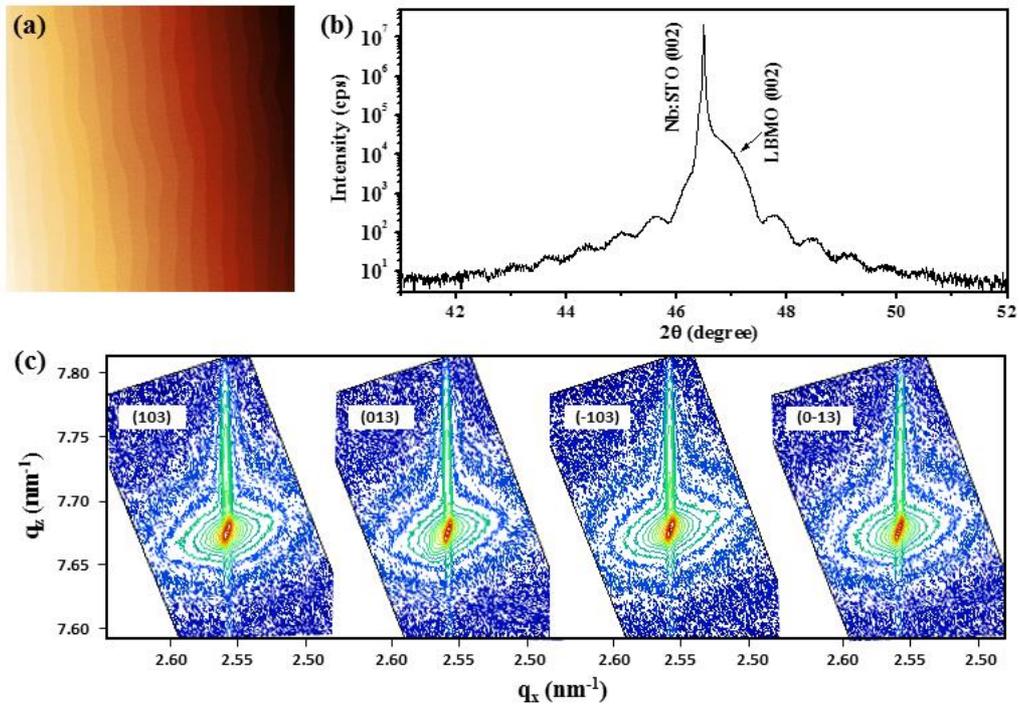

Figure 2. Structural characterization of $La_{0.9}Ba_{0.1}MnO_3$ film grown on $TiO_2$-terminated Nb: $SrTiO_3$ (001) with a thickness of 40 unit cells: (a) A typical $5 \times 5$ $\mu$m AFM height image. (b) XRD θ-2θ detailed scan around (002) reflection. (c) Reciprocal space maps of (103), (013), (10-3), and (0-13) Bragg reflections of Nb: $SrTiO_3$ and $La_{0.9}Ba_{0.1}MnO_3$.

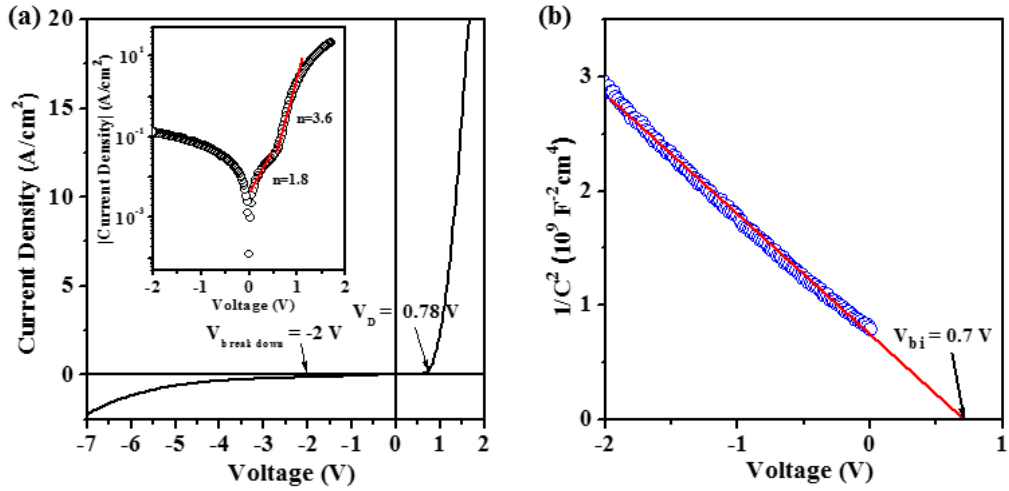

Figure 3. (a) Rectifying current density-voltage (The inset shows the corresponding current density-voltage in semi-logarithmic scale.) and (b) capacitance-voltage characteristics of $La_{0.9}Ba_{0.1}MnO_3/Nb$: $SrTiO_3$ heterojunctions at room temperature for the thickness of 40 unit cells.

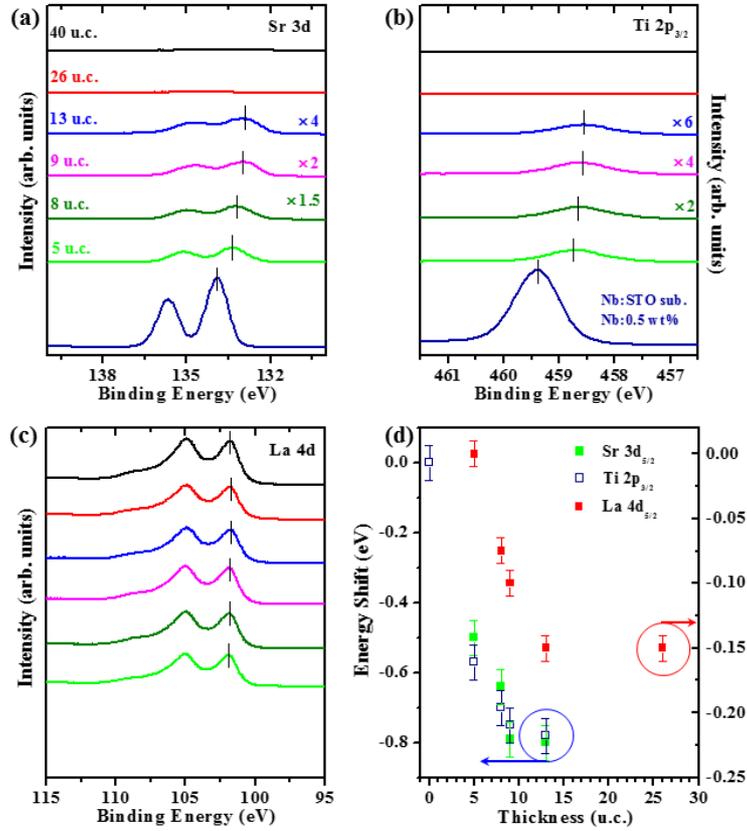

Figure 4. La$_{0.9}$Ba$_{0.1}$MnO$_3$-overlayer-thickness dependence of Sr 3d (a) and Ti 2p$_{3/2}$ (b) core-level spectra of buried TiO$_2$-terminated Nb-doped SrTiO$_3$, (c) La 4d core-level spectrum. (d) Plot of the binding energy shift of the Sr 3d$_{5/2}$, Ti 2p$_{3/2}$, and La 4d$_{5/2}$ core-level peaks as a function of the La$_{0.9}$Ba$_{0.1}$MnO$_3$ overlayer thickness.

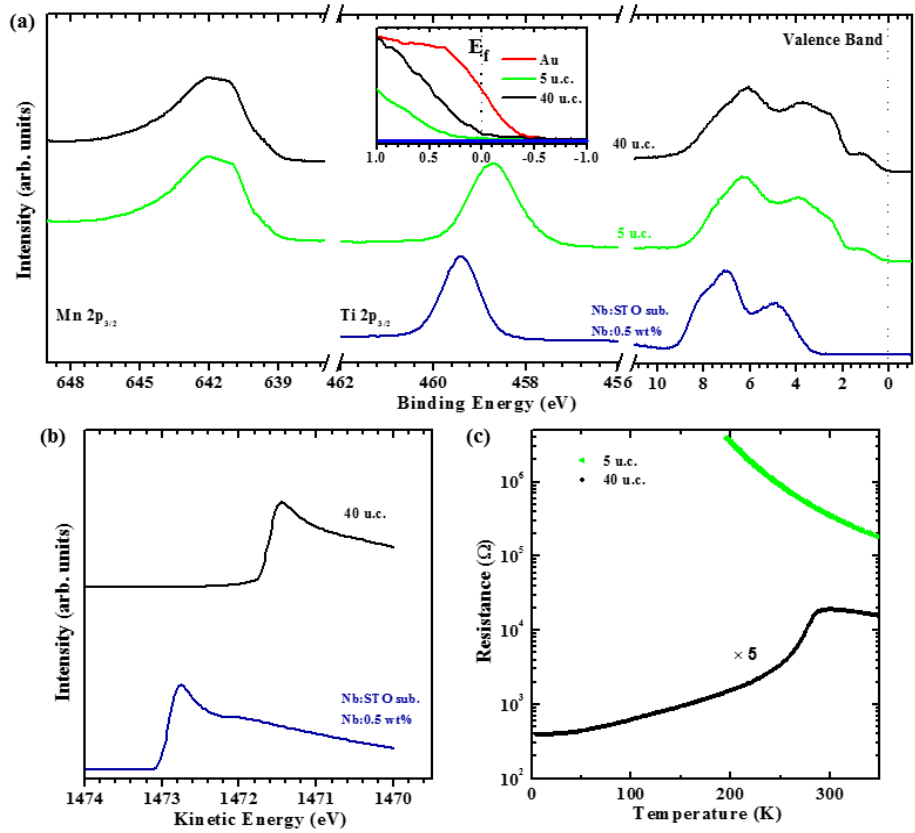

Figure 5. (a) Mn $2p_{3/2}$, Ti 2p $2p_{3/2}$, and XPS valence band spectra for a 40 unit cell La$_{0.9}$Ba$_{0.1}$MnO$_3$ film, a 5 unit cell La$_{0.9}$Ba$_{0.1}$MnO$_3$ film, and a 0.5 wt% Nb-doped SrTiO$_3$ substrate. The inset shows the XPS valence band spectra near the Fermi level (E$_f$), along with the spectrum from an Au foil for the purpose of energy calibration. (b) Secondary energy cutoffs for a 40 unit cell La$_{0.9}$Ba$_{0.1}$MnO$_3$ film and a 0.5 wt% Nb-doped SrTiO$_3$ substrate. (c) Temperature dependence of resistance for 5 unit cell (green) and 40 unit cell (black) La$_{0.9}$Ba$_{0.1}$MnO$_3$ films grown on SrTiO$_3$ substrates.

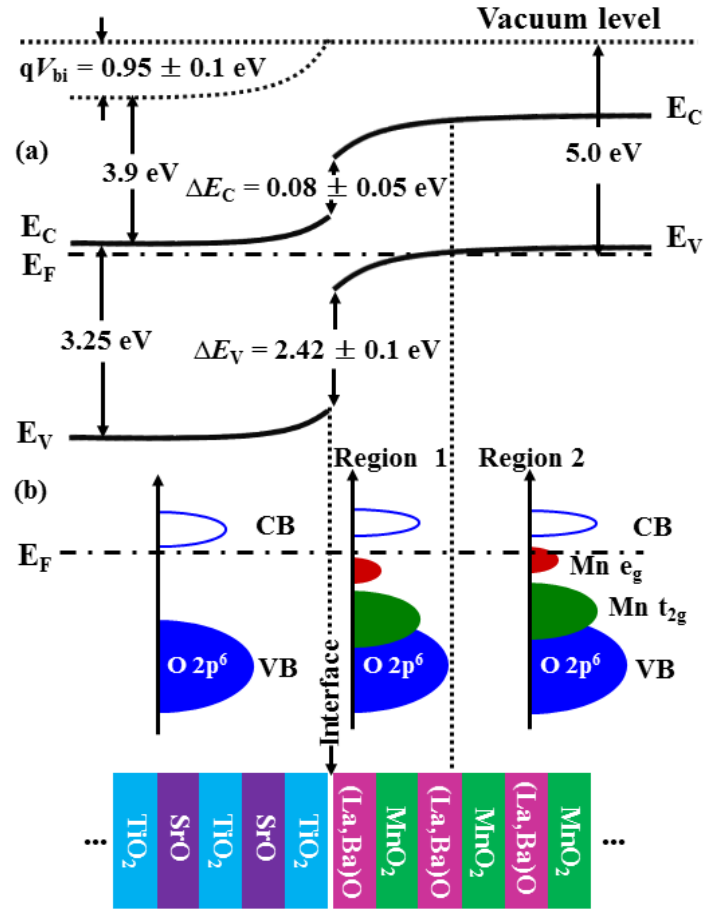

Figure 6. (a) Band diagram and (b) schematic density of states for La$_{0.9}$Ba$_{0.1}$MnO$_3$/Nb:SrTiO$_3$ heterojunctions derived from the present *in-situ* XPS measurements. The bandgap of Nb:SrTiO$_3$ is taken as 3.25 eV. The work functions of the La$_{0.9}$Ba$_{0.1}$MnO$_3$ and the Nb:SrTiO$_3$ substrate are 5.0 eV and 3.9 eV, respectively. The $qV_{bi}$ is built-in potential, the $\Delta E_C$ is conduction band offsets (CBOs), the $\Delta E_V$ is valence band offsets (VBOs), the CB is conduction band, and the VB is valence band.